\documentclass[twocolumn,prl,amsmath,amssymb]{revtex4-1}

\usepackage[pdftex]{graphicx}
\usepackage{dcolumn}
\usepackage{bm}
\usepackage[T1]{fontenc}
\usepackage[utf8]{inputenc}
\usepackage{natbib}
\usepackage{hyperref}
\usepackage{color}
\usepackage{url}

\definecolor{OrangePastel}{RGB}{255,200,31}
\definecolor{GreenPastel}{RGB}{33,219,77}
\definecolor{VioletPastel}{RGB}{200,175,242}
\definecolor{RedPastel}{RGB}{255,125,82}

\definecolor{GreenDarkPastel}{RGB}{33,150,77}
\definecolor{GreenLightPastel}{RGB}{33,255,77}

\definecolor{GreenBabethPastel}{RGB}{209,255,108}
\definecolor{BlueBabethPastel}{RGB}{119,207,255}

{\rule{20pt}{1ex}\endlist}
\let\oldmarginpar\marginpar
\renewcommand\marginpar[1]{\-\oldmarginpar[\raggedleft\footnotesize #1]%
{\raggedright\footnotesize #1}}

\graphicspath{{./../pics/}}

\begin{document}
% New commands
\newcommand{\comment}[1]{
\textcolor{red}{\textbf{\textsf{{\large[}#1{\large]}}}}
}

\title{Large bubble rupture sparks fast liquid jet}
\author{Thomas Séon}%\email{seon@dalembert.upmc.fr}
\affiliation{
Université Pierre et Marie Curie and Centre National de la Recherche 
Scientifique, Unité Mixte de Recherche 7190, Institut Jean Le Rond d’Alembert, 4 
Place Jussieu, F-75005 Paris, France}

\author{Arnaud Antkowiak}%\email[Corresponding author : ]{arnaud.antkowiak@upmc.fr}
\affiliation{
Université Pierre et Marie Curie and Centre National de la Recherche 
Scientifique, Unité Mixte de Recherche 7190, Institut Jean Le Rond d’Alembert, 4 
Place Jussieu, F-75005 Paris, France}

\date{\today}

% Abstract
\begin{abstract}
This Letter presents the novel experimental observation of long and 
narrow jets shooting out in disconnecting large elongated bubbles.
We investigate this phenomenon by carrying out 
experiments with various viscosities, surface tensions, densities and nozzle radii. We propose 
a universal scaling law for the jet velocity, which unexpectedly involves the 
bubble height to the power 3/2. This anomalous exponent suggests an energy focusing phenomenon. We demonstrate experimentally that this focusing is purely gravity-driven and independent of the pinch-off singularity.
\end{abstract}

\maketitle

Disrupting bubbles often exhibit violent jets during their ultimate fast 
out-of-equilibrium dynamics. Examples include bursting 
sea-\citep{Blanchard1967,Duchemin2002} or champagne-bubbles 
\citep{Liger-Belair2009}, collapsing cavitation clouds \citep{Benjamin1966} or 
submarine explosions \citep{Lavrentiev1980} up to the astronomically sized 
buoyant bubble in elliptical galaxy M87 \citep{Churazov2001}.
Among these violent jets, some are the signature of a finite-time singularity \citep{Zeff2000}
 occurring only for particular values of the control parameters \citep{Brenner2000}, but in many cases, including those described here, jet formation is simply the result of a nonsingular relaxation process \citep{Antkowiak2007}.

Before bursting however, bubbles are not known to exhibit interface deformation as intense as the ones just depicted. Indeed, 
irrespective of their initial distortion \citep{Yang2003} small bubbles gently 
relax toward robust equilibrium shapes in the form of spheres or oblate spheroids 
\citep{Magnaudet2000}. As regards the final form of larger bubbles, initial shape 
matters. According to their initial appearance, weakly deformed bubbles smoothly evolve either to spherical caps 
\citep{Davies1950} or toroidal bubbles \citep{Walters1963, Bonometti2006}, as the 
ones typically expelled by dolphins or divers \citep{Marten1996}. Much less is known about 
initially strongly deformed bubbles, as their dynamics and outcome have barely been studied. Such bubbles are yet commonplace in bubbling systems typically 
found in geophysics or industry (\textit{e.g.} occurring in glassy 
\citep{Shelby2005} or metallic \citep{Guezennec2005} melts). The primary aim of the present study is therefore to investigate the transient behavior of such large and highly deformed bubbles, which will prove to be as intense as bursting events. 

\begin{figure*}[htd]
\noindent\includegraphics[width=\hsize]{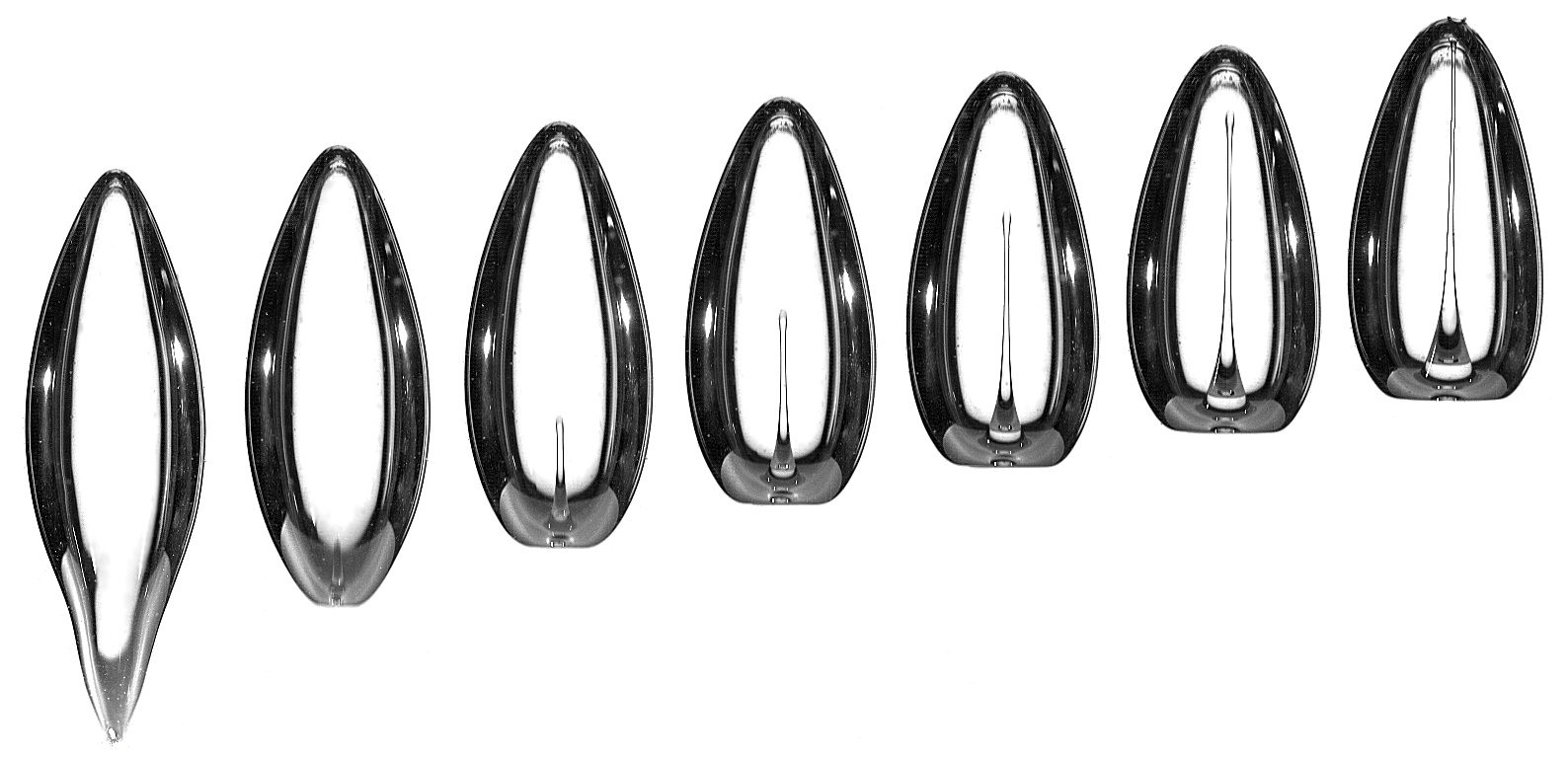}
\caption{Time sequence of the jet developing inside a large bubble just after its 
release from a submerged nozzle. The large bubble pinches off at the first image 
and the back-to-equilibrium dynamics exhibits an intense and concentrated jet. 
The time lapse between the snapshots is $\Delta t$ = 3 ms. The bubble height, 
right after detachment, is $H$ = 4.22 cm, the bubble front velocity is 
$V_\text{front}$ = 0.67~m.s$^{-1}$ and the jet velocity is $V_\text{jet}$ = 
2.69~m.s$^{-1}$. The liquid viscosity is $\mu = 420$ mPa.s. The gas is injected 
through an injector of diameter $d = 1.8$ mm at an airflow rate of $Q=$ 4.4 
$\ell$.min$^{-1}$. The fluid height is $h= 10$ cm.} 
\label{fig:sequence_jet}
\end{figure*}

In this Letter, we report a surprising violent jet dynamics following large 
bubble disconnection (Fig.~\ref{fig:sequence_jet}) in a bubbling experiment. 
The thin and concentrated jet developing inside the bubble possibly gives rise to 
liquid projections shooting out way above the free surface. We analyze 
experimentally this phenomenon and we provide evidences of the role played by 
gravity in the jet formation.

% Figure 1

Our experiments consist in releasing gas bubbles from a submerged orifice in a 
viscous liquid. The liquid is contained in a transparent tank taken sufficiently 
large  (20 cm $\times$ 20 cm $\times$ 25 cm) to rule out confinement effects. It 
was observed that liquid slugs in the feeding line could make the bubble 
detachment frequency strongly fluctuate. Those slugs originate in a liquid 
invasion, either caused by capillarity or gravity. Capillary invasion is 
circumvented by the use of non-wetting aqueous liquids instead of oils. Therefore 
the liquids used in this study include sugar cane syrup (Canadou) of viscosity 
$\mu=110$ mPa.s and surface tension $\gamma = 90$ mN.m$^{-1}$ and three 
water-glycerol mixtures of viscosity $\mu = 140$, $280$, $420$ mPa.s and  surface 
tension $\gamma = 65$ mN.m$^{-1}$. In order to both allow the development of the 
bubbles and mitigate gravity-driven invasion (high hydrostatic pressure on the 
orifice), the height $h$ of fluid above the injector is typically kept in the 
range 5-10 cm. Air is fed through a nozzle of diameter $1.8$ mm. Injection is 
controlled by a mass flow meter (Alicat Scientific) that provides a constant 
flowrate $Q$ by adjusting the air pressure. It allows us to achieve a wide range 
of airflows (from 0.01 to 10 $\ell$.min$^{-1}$). The bubble and jet dynamics are 
analyzed through ultra-fast imagery. To do so, the tank is back lit and images 
are obtained at $4000$ frames per second using a digital high-speed camera 
(Photron SA-5). 

Figure~\ref{fig:sequence_jet} illustrates a typical jetting event following the 
disconnection of a large gas bubble. The entire sequence lasts $18$ ms and the 
bubble height, right after detachment, is $H = 4.22$ cm. Bubble pinch-off occurs 
on the first image and instantly a narrow, high-speed vertical jet shoots out 
inside the bubble. It is noteworthy that the bubble shape hardly deforms 
throughout the development of the jet, except near the fast recoiling conical rear. As the jet velocity $V_\text{jet}$ = 
2.69~m.s$^{-1}$ is much higher than the bubble front velocity $V_\text{front}$ = 
0.67~m.s$^{-1}$, the jet tip reaches the top of the bubble in a few milliseconds. 
Here, the jet is suddenly stopped as it collides with the bubble wall, but in 
other cases the jet literally perforates the bubble and makes headway in the bulk 
liquid. 

\begin{figure}
%\noindent\includegraphics{figure2/figure2_light.pdf}
\noindent\includegraphics{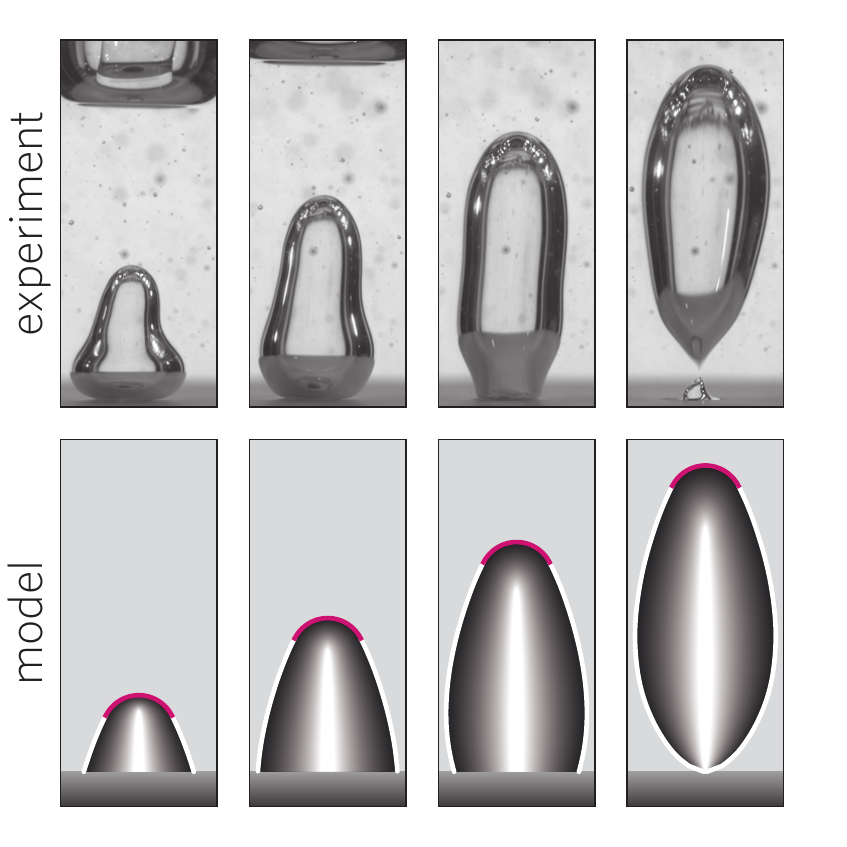}
\caption{Comparison of experimental and simulated time sequences of bubble 
formation. In the 2D Rayleigh-Plesset model, the frozen spherical cap is 
materialized by a thick red line. The time lapse between each snapshot in the 
model and in the experiment is $\Delta t~=~11$~ms. The physical parameters are 
the air flowrate $Q = 3.80~\ell.$min$^{-1}$, liquid viscosity $\mu = 280$~mPa.s, 
density $\rho = 1250$ kg.m$^{-3}$ and surface tension $\gamma = 65$~mN.m$^{-1}$. 
The parameters of the model $R_\text{front} = 4.18$ mm and $V_\text{front} = 
0.785$~m.s$^{-1}$ are given by the experimental data. $R_\infty =~18.3$~mm  and $
\theta = 1.1$ rad are adjusted to the dynamics of the experiment. The slight 
shape dissimilarity is related to the absence of vertical motions in the 
theoretical model. Nonetheless, the seal timescale and overall dynamics are in 
fair agreement.} 
\label{fig:comp_modele}
\end{figure}

In addition to the classical forces determining bubble volume 
\citep{Clift1978}, the detachment of the bubble is here potentially influenced by 
several other complex processes, including viscous stresses associated with the 
large scale convection pattern, gas pressure at the inlet, wake of the foregoing 
bubble\ldots\ However we argue in the following that detachment is primarily due 
to the gravity-driven collapse of the bubble neck. We derive a simple model for 
the bubble formation, considering principally the expanding and contracting 
liquid motions induced by the bubble growth and seal process. The ambient liquid 
is mimicked by a finite set of slices in pure radial motion around a pressurized 
cavity \citep{Lohse2004,Duclaux2007}. Assuming a potential flow evolution, we can obtain an equation for the 
position of the free boundary~$R(t)$ for each slice:
\begin{equation*}
\left(R\ddot{R}+\dot{R}^2\right) \ln \frac{R}{R_\infty} + \frac{\lambda}{2} \dot
{R}^2 
= -\frac{P_\text{bubble}}{\rho} - g z + \frac{2 \mu \dot R + 
\gamma}{\rho R}.
\end{equation*}
This is the 2D version of Rayleigh-Plesset equation describing the liquid motion 
around a hollow cylinder \citep{Oguz1993,Burton2005}, with $\rho$ the liquid 
density, $g$ the gravity, $\lambda =  1 - R^2/R_\infty^2$ a confinement factor,
$P_\text{bubble} = -\rho g H(t) + \gamma \kappa$ the bubble pressure (taking the reference level at $z=0$), 
$\kappa = 2/R_\text{front}$ the front curvature, $R_\infty$ a distant cut-off, 
and $H(t)=V_\text{front}\, t$ and $z$ denoting respectively the height of the 
bubble and the local position of the slice. The bubble front is approximated by a 
frozen spherical cap of angle $2 \theta$ rising at constant velocity. The cavity 
formation results from the air entrained behind it. The initialization procedure 
for each slice below this traveling bubble front simulates the deflection of 
fluid particles past it. At the trailing edge of the sphere portion, 
\textit{i.e.} at the point $R(t) = R_\text{front} \sin \theta$ the following 
velocity is imposed: $\dot R(t) = \mathrm dR/\mathrm dt=V \cot \theta$.
Figure~\ref{fig:comp_modele} shows a typical face-to-face comparison between the 
experiment and the model. Remarkably, the overall detachment sequence is well 
captured even if the axial motions are disregarded \footnote{In the context of 
the deep seal of a transient impact-generated cavity, the 2D Rayleigh-Plesset 
model is also in excellent agreement with the experiments, even if the vertical 
motions are neglected \citep{Bergmann2009}. This suggests that the relevant 
mechanism for the pinch-off is radial in essence.}. This agreement between the 
experiment and a rough model containing as only driving force hydrostatic 
pressure constitutes a strong evidence that the seal mechanism is gravity-driven.

% Figure 2

% Figure 3
\begin{figure}[td]
\noindent
\includegraphics{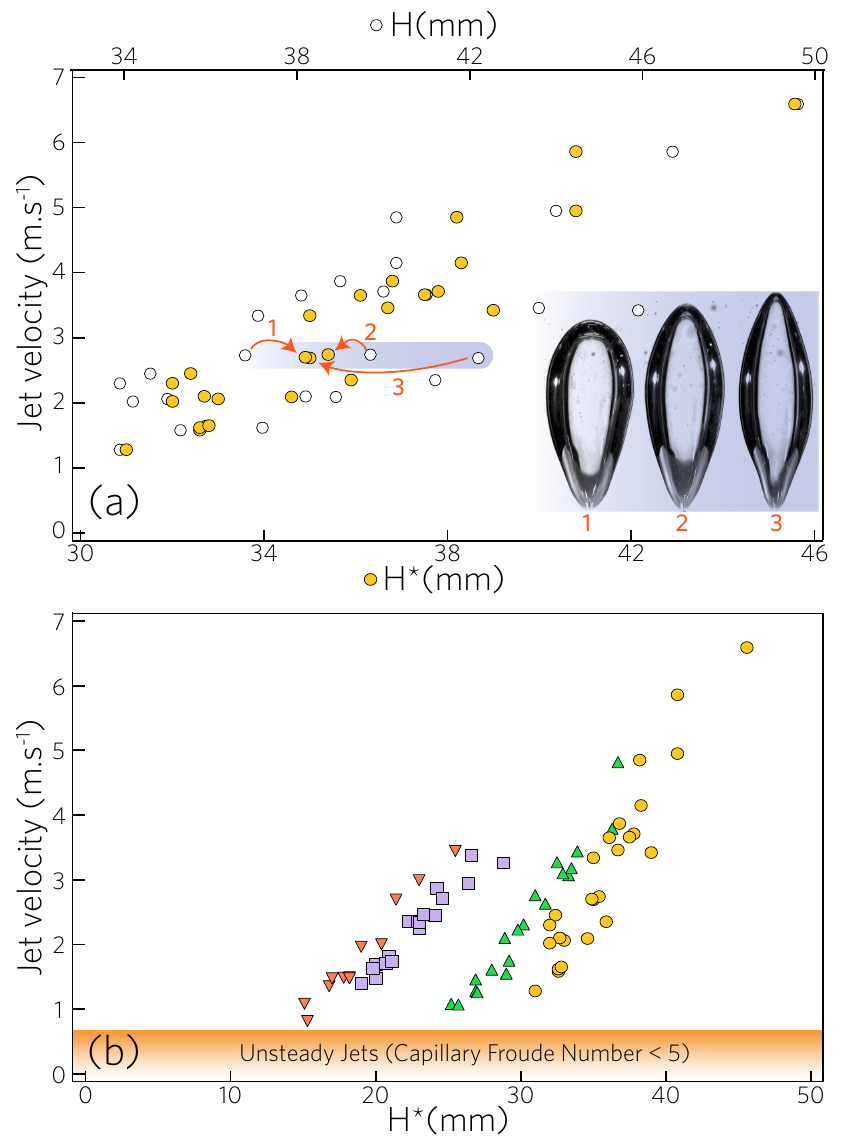}
\caption{(a) Jet tip velocity ($V_\text{jet}$) as a function of heights: 
$H$~($\circ$) height of the bubble after detachment (top axis) and 
$H^{\star} = H - 2\gamma/\rho g R_\text{front}$ (\textcolor{OrangePastel}{
$\bullet$}) taking into account the capillary pressure term (bottom axis). The 
data spread as a function of $H$ significantly decreases when plotted as a 
function of $H^{\star}$. The three bubbles presented in the bottom right corner 
labelled 1, 2 and 3 have the same jet velocity, but for a different height $H$ 
(highlighted in blue). When plotted with respect to $H^\star$, they all gather. 
The viscosity of the liquid is $\mu=420$ mPa.s. (b) $V_\text{jet}$ as a function 
of  $H^{\star}$ for four different viscosities $\mu$~: $110$ mPa.s 
(\textcolor{RedPastel}{$\blacktriangledown$}),  $140$ mPa.s 
(\textcolor{VioletPastel}{$\blacksquare$}),  $280$ mPa.s  
(\textcolor{GreenPastel}{$\blacktriangle$}),  $420$ mPa.s 
(\textcolor{OrangePastel}{$\bullet$}). The surface tension and density are 
$\gamma = 65$ mN.m$^{-1}$ and $\rho = 1250$ kg.m$^{-3}$ except for the lowest 
viscosity (\textcolor{RedPastel}{$\blacktriangledown$}) for which $\gamma = 90$ 
mN.m$^{-1}$ and $\rho = 1350$ kg.m$^{-3}$. The injector diameter is $d=1.8$ mm and 
the fluid height is $h= 10$ cm.} 
\label{fig:velocityvsH}
\end{figure}

We now turn to the jet formation per se, occurring right after the cavity closing. 
Fig.~\ref{fig:velocityvsH}a shows the 
experimental relation between the jet velocity $V_\text{jet}$ and the bubble 
height $H$, right after pinch-off instant. Each circle ($\circ$) corresponds to a 
given experiment conducted with fixed ambient viscosity $\mu = 420$ mPa.s, 
injector diameter $d = 1.8$ mm and fluid height $h = 10$ cm (same physical 
parameters as in Fig.~\ref{fig:sequence_jet}). For all experiments, it was found 
from spatiotemporal diagram analysis that the velocity of the jet rapidly reaches 
a constant value after a transient regime. A clear trend appears from this 
figure: the jet velocity $V_\text{jet}$ increases with the bubble height $H$, 
consistently with a gravity-powered mechanism. Though exhibiting a visible 
tendency, the data are still quite spread out as for one value of the jet 
velocity correspond several bubble heights. To understand this spreading, let us 
examine the initial shape of some typical bubbles for a specific value 
$V_\text{jet}\simeq2.7$~m.s$^{-1}$. In Fig.~\ref{fig:velocityvsH}(a) are 
represented three bubbles, marked from 1 to 3, corresponding to the highlighted 
circles (bubble number 3 is the same as in Fig.~\ref{fig:sequence_jet}). We 
observe that the three bubbles have a completely different shape, presenting a 
wide range of front curvature. Interestingly, the largest bubble (3) also bears 
the highest front curvature, while the smallest bubble (1) is the less curved. 
Yet in spite of these differences their jet velocities are the same. This 
geometrical competition between bubble height and front radius of curvature 
suggests a physical competition between an hydrostatic driving force and a 
capillary quenching effect: the main source of pressure difference between the 
liquid and the bubble is gravity, and capillarity tends to lessen this difference 
by pressurizing the bubble. In order to take both effects into account, we 
represent in Fig.~\ref{fig:velocityvsH}(a) the jet velocity $V_\text{jet}$ versus 
the capillary corrected bubble height~$H^{\star} = H - 2\gamma/\rho g R_\text{front}$; $H^\star$~is a natural lengthscale in the sense that the bubble pressure is approximately $ -\rho g H^\star$ 
(see orange circles (\textcolor{OrangePastel}{$\bullet$}) related to the bottom 
axis). Upon using $H^\star$ instead of $H$, the three tracked experiments now get 
together as indicated by the arrows in Fig.~\ref{fig:velocityvsH}(a). A similar 
gathering is observed for all experiments. This indicates the relevance of the 
parameter $H^\star$ and points to a pressure-driven mechanism for the jet 
formation, as is common in impact-driven \citep{Antkowiak2007} or gravity-driven 
jets \citep{Bergmann2008}.

The influence of viscosity $\mu$ and surface tension $\gamma$ in the development 
of the jet has been investigated as well by conducting similar experiments in 
various liquids. The results are summarized in Fig.~\ref{fig:velocityvsH}(b). For 
each viscosity the same rise in jet velocity $V_\text{jet}$ with $H^\star$ is 
observed, albeit with a general shift to higher $H^\star$ with more viscous 
liquids. It is worth noting that jets still exist for values of the control 
parameters corresponding to the shaded area in Fig.~\ref{fig:velocityvsH}(b). But 
in this region, the jet velocity never reaches a constant value. Such unsteady 
jets were systematically disregarded during postprocessing. This shaded area 
corresponds to a range of `capillary Froude number' 
$V_\text{jet}/\sqrt{\gamma/\rho r}$ below a typical value of 5. There, the
 feeding velocity $V_\text{jet}$ is too close to the retraction velocity $\sim 
\sqrt{\gamma/\rho r}$, with $r$ the jet radius, hence the unsteadiness.  

\begin{figure}[ht]
\noindent\includegraphics{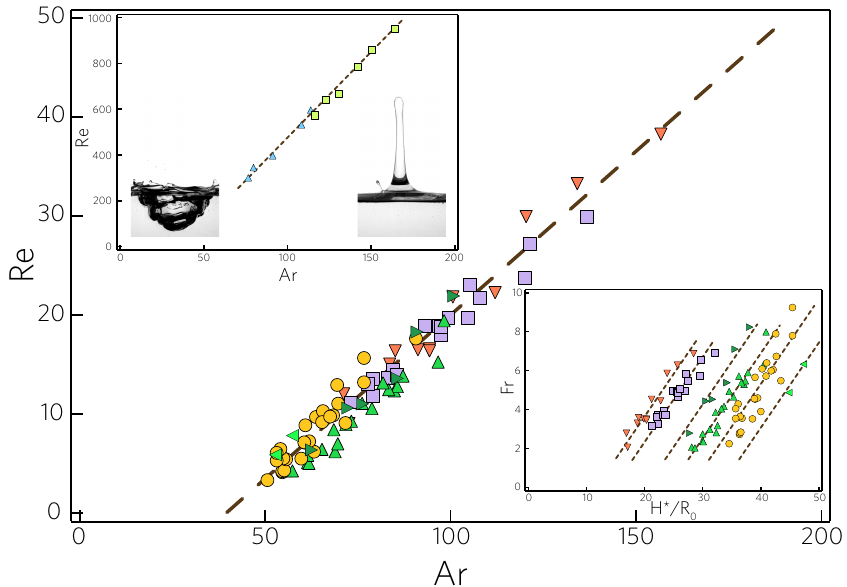}
\caption{Jet Reynolds number Re = $\rho V_\text{jet} R_o / \mu$ versus Archimedes 
number Ar = $ \rho \sqrt{gH^{\star}} H^{\star}/\mu$ for different viscosities, 
surface tensions, densities and nozzle radii. 
The significance of the symbols is the same as in 
Fig. \ref{fig:velocityvsH}(b) and two series were added corresponding to two different injector diameters : 2.15 mm (\textcolor{GreenDarkPastel}{$\blacktriangleleft$}) and 1.19 mm (\textcolor{GreenLightPastel}{$\blacktriangleright$}) with viscosity, surface tension and density respectively $\mu=320$ mPa.s, $\gamma = 65$ mN.m$^{-1}$ and $\rho = 1250$ kg.m$^{-3}$.
The equation of the oblique dashed line is Re = 
$\alpha$ Ar - $\beta$, with $\alpha = 0.33$ and $\beta = 13$. The bottom right inset shows the 
Froude number Fr = $V_\text{jet}/\sqrt{g H^\star}$ versus the nondimensional 
bubble height $H^\star/R_o$. The lines act as eye guides and present the same slope 
$\alpha$. The top left inset presents the results of a different experiment where the jet is formed due to the gravitational collapse of a free surface hollow.  By taking $H$ the depth of the cavity and $L$ the maximum
cavity diameter as relevant scales, jet Reynolds number Re = $\rho V_\text{jet} L / \mu$ is plotted  versus Archimedes 
number Ar = $ \rho \sqrt{gH} H/\mu$ for two different viscosities 
$\mu$~: $220$ mPa.s (\textcolor{GreenBabethPastel}{$\blacksquare$}) and $426$ mPa.s 
(\textcolor{BlueBabethPastel}{$\blacktriangle$}). The two snapshots display a typical cavity ($H = 5.61$~cm and $L= 12.6$~cm) and the resulting jet.} 
\label{fig:rescaling}
\end{figure}

Now taking the natural gravito-inertial velocity $\sqrt{g H^\star}$ and the 
nozzle radius $R_o$ as relevant scales of the problem, we non-dimensionalize our 
results. The rescaled jet velocity or Froude number Fr = 
$V_\text{jet}/\sqrt{g H^{\star}}$ is plotted against the dimensionless height 
$H^{\star}/R_o$ in the bottom right inset of Fig.~\ref{fig:rescaling} for all viscosities,
surface tensions and nozzle radii. Surprisingly enough, the six corresponding curves increase 
with respect to the dimensionless height. Actually we might have expected from a 
balance between buoyancy and inertia to observe a constant Froude number 
\citep{Bergmann2008}. Rather we observe a linear variation of the Froude number 
with $H^{\star}/R_o$, identical for all sets of experiments. Moreover a shift to 
lower non-dimensional velocities is noticeable as the viscosity is increased. 
This points to a functional dependence of the Froude number with $H^{\star}/R_o$ 
and the dimensionless viscosity as follows:
\begin{equation*}
\frac{V_\text{jet}}{\sqrt{g H^\star}} = \alpha \frac{H^\star}{R_o} - \mathcal{F}
\left(\frac{\mu}{\rho \sqrt{g H^\star} R_o}\right),
\end{equation*}
with $\alpha$ a non-dimensional constant. Analysis of the experimental data 
reveals that the form of function $\mathcal F\left(x\right)$ is actually a linear 
function $\beta x$, with $\beta$ constant. This allows 
us to rewrite the preceding relation with a single identical offset for all 
viscosities:
\begin{equation*}
\text{Re} = \alpha \,\text{Ar} - \beta,
\end{equation*}
where we have introduced the jet Reynolds number Re = $\rho V_\text{jet} R_o/\mu$
and Archimedes number Ar = $\rho \sqrt{g H^\star} H^\star/\mu$.
%--- FIG 4

Figure~\ref{fig:rescaling} represents the relation between the jet Reynolds 
number and Archimedes number for all the conducted experiments. The collapse of 
the six series of experiments corresponding to five different viscosities, 
two surface tensions and densities and three different nozzle diameters is excellent. The relation between the jet 
velocity and the capillary corrected height of the bubble for the whole range of 
physical parameters is finally captured with the simple law Re = $\alpha$ Ar - $
\beta$, taking $\alpha = 0.33$ and $\beta = 13$. In systems exhibiting a balance between buoyancy and inertia, the potential volume energy $\sim \rho g H$ is converted into kinetic volume energy $\sim \rho V^2$ and the velocity typically scales as $H^{1/2}$: large 
Taylor bubbles in tubes \citep{Davies1950}, gravity waves in shallow fluid layers 
\citep{Whitham1974}, inertial gravity current \citep{Simpson1997}\ldots\
Surprisingly, the here obtained scaling law shows a dependence of the jet velocity $V_\text{jet}$ with 
${H^{\star}}^{3/2}$ rather than ${H^{\star}}^{1/2}$. This anomalous exponent suggests an energy focusing phenomenon whose source is now discussed. 

Does the pinch-off singularity play a role in this focusing? In order to investigate the role of detachment, we have designed and carried out a new model experiment, free of pinch-off. This experiment consists in blowing air over a free liquid surface so as to form a depression, mimicking the conical rear of the bubble. Upon the collapse of this hollow, a gravity-driven jet develops (see snapshots fig.~\ref{fig:rescaling}). But there again, the same scaling for the jet velocity with $H^{3/2}$ is observed (top left inset). Note also that, in sharp contrast with singular focusing behaviors \citep{Zeff2000}, the cavity reversal observed here is not unlike the shell eversion process in that jet eruption/curvature reversal happens  before collapse.  These rule out both pinch-off \citep{Bolanos-Jimenez2008} and curvature singularities \citep{Zeff2000} in the observed energy focusing and demonstrate that it is merely a signature of the gravitational cavity collapse.

In conclusion, the present experimental work reports on the violent dynamics exhibited during the relaxation of an initially large oblate bubble. The intense and narrow jets developing inside the detaching bubbles follow a surprising dependence with ${H^\star}^{3/2}$. The related focusing of energy is not a consequence of the detachment singularity, as proven by the persistence of the scaling in a pinch-off-free setting. Consequently, in this experiment at least, the jet does not keep the footprint of the singularity as in \textit{e.g.} \citet{Zeff2000}, \citet{Bartolo2006} or \citet{Gekle2010}. Instead, global conservation rules take over the jet dynamics. The scaling law also suggests the existence range of this type of jet, with a threshold in Archimedes given by $\beta/\alpha$ ($\simeq 40$ in our 
case). Below this value, the deformation of the bubble rather tends to the unsteady liquid tongue of \citet{Walters1963}. Alternatively, the threshold defines a critical 
viscosity-dependent bubble height necessary to observe those liquid jets. Finally, in this paper we have focused on the 
jet development before its collision with the bubble front. An open remaining 
question is whether the jet could have a sufficient energy to perforate the 
bubble and reach the free surface. In our experiments we observed that strong 
liquid projections can appear above the free surface for large airflow rate. We 
are currently investigating the highly unclear link between the jet presented 
here and these liquid projections.

% Acknowledgments
\begin{acknowledgments}
L’Agence Nationale de la Recherche through its Grant “DEFORMATION” 
ANR-09-JCJC-0022-01 and the Émergence(s) program of the Ville de Paris are 
acknowledged for their financial support. AA is particularly grateful to Virginie 
Duclaux for pointing out the here presented phenomenon. We also thank José Manuel 
Gordillo for stimulating discussions on bubble pinch-off subtleties.
\end{acknowledgments}

% Bibliography
\bibliographystyle{apsrev4-1}
\bibliography{BiblioPRL}

%merlin.mbs 2010-03-15 4.21a (PWD, AO, DPC)
\begin{thebibliography}{30}%
\makeatletter
\providecommand \@ifxundefined [1]{%
 \@ifx{#1\undefined}
}%
\providecommand \@ifnum [1]{%
 \ifnum #1\expandafter \@firstoftwo
 \else \expandafter \@secondoftwo
 \fi
}%
\providecommand \@ifx [1]{%
 \ifx #1\expandafter \@firstoftwo
 \else \expandafter \@secondoftwo
 \fi
}%
\providecommand \natexlab [1]{#1}%
\providecommand \enquote  [1]{``#1''}%
\providecommand \bibnamefont  [1]{#1}%
\providecommand \bibfnamefont [1]{#1}%
\providecommand \citenamefont [1]{#1}%
\providecommand \href@noop [0]{\@secondoftwo}%
\providecommand \href [0]{\begingroup \@sanitize@url \@href}%
\providecommand \@href[1]{\@@startlink{#1}\@@href}%
\providecommand \@@href[1]{\endgroup#1\@@endlink}%
\providecommand \@sanitize@url [0]{\catcode `\\12\catcode `\$12\catcode
  `\&12\catcode `\#12\catcode `\^12\catcode `\_12\catcode `\%12\relax}%
\providecommand \@@startlink[1]{}%
\providecommand \@@endlink[0]{}%
\providecommand \url  [0]{\begingroup\@sanitize@url \@url }%
\providecommand \@url [1]{\endgroup\@href {#1}{\urlprefix }}%
\providecommand \urlprefix  [0]{URL }%
\providecommand \Eprint [0]{\href }%
\@ifxundefined \urlstyle {%
  \providecommand \doi  [0]{\begingroup \@sanitize@url \@doi}%
  \providecommand \@doi [1]{\endgroup \@@startlink {\doibase
  #1}doi:\discretionary {}{}{}#1\@@endlink }%
}{%
  \providecommand \doi  [0]{doi:\discretionary{}{}{}\begingroup
  \urlstyle{rm}\Url }%
}%
\providecommand \doibase [0]{http://dx.doi.org/}%
\providecommand \Doi [0]{\begingroup \@sanitize@url \@Doi }%
\providecommand \@Doi  [1]{\endgroup\@@startlink{\doibase#1}\@@Doi}%
\providecommand \@@Doi [1]{#1\@@endlink}%
\providecommand \selectlanguage [0]{\@gobble}%
\providecommand \bibinfo  [0]{\@secondoftwo}%
\providecommand \bibfield  [0]{\@secondoftwo}%
\providecommand \translation [1]{[#1]}%
\providecommand \BibitemOpen [0]{}%
\providecommand \bibitemStop [0]{}%
\providecommand \bibitemNoStop [0]{.\EOS\space}%
\providecommand \EOS [0]{\spacefactor3000\relax}%
\providecommand \BibitemShut  [1]{\csname bibitem#1\endcsname}%
%</preamble>
\bibitem [{\citenamefont {Blanchard}(1967)}]{Blanchard1967}%
  \BibitemOpen
  \bibfield  {author} {\bibinfo {author} {\bibfnamefont {D.~C.}\ \bibnamefont
  {Blanchard}},\ }\href@noop {} {\emph {\bibinfo {title} {From raindrops to
  volcanoes}}}\ (\bibinfo  {publisher} {Doubleday},\ \bibinfo {year}
  {1967})\BibitemShut {NoStop}%
\bibitem [{\citenamefont {Duchemin}\ \emph {et~al.}(2002)\citenamefont
  {Duchemin}, \citenamefont {Popinet}, \citenamefont {Josserand},\ and\
  \citenamefont {Zaleski}}]{Duchemin2002}%
  \BibitemOpen
  \bibfield  {author} {\bibinfo {author} {\bibfnamefont {L.}~\bibnamefont
  {Duchemin}}, \bibinfo {author} {\bibfnamefont {S.}~\bibnamefont {Popinet}},
  \bibinfo {author} {\bibfnamefont {C.}~\bibnamefont {Josserand}}, \ and\
  \bibinfo {author} {\bibfnamefont {S.}~\bibnamefont {Zaleski}},\ }\href@noop
  {} {\bibfield  {journal} {\bibinfo  {journal} {Phys. Fluids},\ }\textbf
  {\bibinfo {volume} {14}},\ \bibinfo {pages} {3000} (\bibinfo {year}
  {2002})}\BibitemShut {NoStop}%
\bibitem [{\citenamefont {Liger-Belair}\ \emph {et~al.}(2009)\citenamefont
  {Liger-Belair}, \citenamefont {Cilindre}, \citenamefont {Gougeon},
  \citenamefont {Lucio}, \citenamefont {Gebef{\"u}gi}, \citenamefont
  {Jeandet},\ and\ \citenamefont {Schmitt-Kopplin}}]{Liger-Belair2009}%
  \BibitemOpen
  \bibfield  {author} {\bibinfo {author} {\bibfnamefont {G.}~\bibnamefont
  {Liger-Belair}}, \bibinfo {author} {\bibfnamefont {C.}~\bibnamefont
  {Cilindre}}, \bibinfo {author} {\bibfnamefont {R.~D.}\ \bibnamefont
  {Gougeon}}, \bibinfo {author} {\bibfnamefont {M.}~\bibnamefont {Lucio}},
  \bibinfo {author} {\bibfnamefont {I.}~\bibnamefont {Gebef{\"u}gi}}, \bibinfo
  {author} {\bibfnamefont {P.}~\bibnamefont {Jeandet}}, \ and\ \bibinfo
  {author} {\bibfnamefont {P.}~\bibnamefont {Schmitt-Kopplin}},\ }\href@noop {}
  {\bibfield  {journal} {\bibinfo  {journal} {Proc. Natl Acad. Sci. USA},\
  }\textbf {\bibinfo {volume} {106}},\ \bibinfo {pages} {16545} (\bibinfo
  {year} {2009})}\BibitemShut {NoStop}%
\bibitem [{\citenamefont {Benjamin}\ and\ \citenamefont
  {Ellis}(1966)}]{Benjamin1966}%
  \BibitemOpen
  \bibfield  {author} {\bibinfo {author} {\bibfnamefont {T.~B.}\ \bibnamefont
  {Benjamin}}\ and\ \bibinfo {author} {\bibfnamefont {A.~T.}\ \bibnamefont
  {Ellis}},\ }\href@noop {} {\bibfield  {journal} {\bibinfo  {journal} {Philos.
  Trans. R. Soc. London, Ser. A},\ }\textbf {\bibinfo {volume} {260}},\
  \bibinfo {pages} {221} (\bibinfo {year} {1966})}\BibitemShut {NoStop}%
\bibitem [{\citenamefont {Lavrentiev}\ and\ \citenamefont
  {Chabat}(1980)}]{Lavrentiev1980}%
  \BibitemOpen
  \bibfield  {author} {\bibinfo {author} {\bibfnamefont {M.}~\bibnamefont
  {Lavrentiev}}\ and\ \bibinfo {author} {\bibfnamefont {B.}~\bibnamefont
  {Chabat}},\ }\href@noop {} {\emph {\bibinfo {title} {Effets hydrodynamiques
  et mod\`eles math\'ematiques}}}\ (\bibinfo  {publisher} {\'Editions MIR},\
  \bibinfo {year} {1980})\BibitemShut {NoStop}%
\bibitem [{\citenamefont {Churazov}\ \emph {et~al.}(2001)\citenamefont
  {Churazov}, \citenamefont {Br{\"u}ggen}, \citenamefont {Kaiser},
  \citenamefont {B{\"o}hringer},\ and\ \citenamefont {Forman}}]{Churazov2001}%
  \BibitemOpen
  \bibfield  {author} {\bibinfo {author} {\bibfnamefont {E.}~\bibnamefont
  {Churazov}}, \bibinfo {author} {\bibfnamefont {M.}~\bibnamefont
  {Br{\"u}ggen}}, \bibinfo {author} {\bibfnamefont {C.~R.}\ \bibnamefont
  {Kaiser}}, \bibinfo {author} {\bibfnamefont {H.}~\bibnamefont
  {B{\"o}hringer}}, \ and\ \bibinfo {author} {\bibfnamefont {W.}~\bibnamefont
  {Forman}},\ }\href@noop {} {\bibfield  {journal} {\bibinfo  {journal}
  {Astrophys. J.},\ }\textbf {\bibinfo {volume} {554}},\ \bibinfo {pages} {261}
  (\bibinfo {year} {2001})}\BibitemShut {NoStop}%
\bibitem [{\citenamefont {Zeff}\ \emph {et~al.}(2000)\citenamefont {Zeff},
  \citenamefont {Kleber}, \citenamefont {Fineberg},\ and\ \citenamefont
  {Lathrop}}]{Zeff2000}%
  \BibitemOpen
  \bibfield  {author} {\bibinfo {author} {\bibfnamefont {B.~W.}\ \bibnamefont
  {Zeff}}, \bibinfo {author} {\bibfnamefont {B.}~\bibnamefont {Kleber}},
  \bibinfo {author} {\bibfnamefont {J.}~\bibnamefont {Fineberg}}, \ and\
  \bibinfo {author} {\bibfnamefont {D.~P.}\ \bibnamefont {Lathrop}},\
  }\href@noop {} {\bibfield  {journal} {\bibinfo  {journal} {Nature},\ }\textbf
  {\bibinfo {volume} {403}},\ \bibinfo {pages} {401} (\bibinfo {year}
  {2000})}\BibitemShut {NoStop}%
\bibitem [{\citenamefont {Brenner}(2000)}]{Brenner2000}%
  \BibitemOpen
  \bibfield  {author} {\bibinfo {author} {\bibfnamefont {M.~P.}\ \bibnamefont
  {Brenner}},\ }\href@noop {} {\bibfield  {journal} {\bibinfo  {journal}
  {Nature},\ }\textbf {\bibinfo {volume} {403}},\ \bibinfo {pages} {377}
  (\bibinfo {year} {2000})}\BibitemShut {NoStop}%
\bibitem [{\citenamefont {Antkowiak}\ \emph {et~al.}(2007)\citenamefont
  {Antkowiak}, \citenamefont {Bremond}, \citenamefont {Le~Diz{\`e}s},\ and\
  \citenamefont {Villermaux}}]{Antkowiak2007}%
  \BibitemOpen
  \bibfield  {author} {\bibinfo {author} {\bibfnamefont {A.}~\bibnamefont
  {Antkowiak}}, \bibinfo {author} {\bibfnamefont {N.}~\bibnamefont {Bremond}},
  \bibinfo {author} {\bibfnamefont {S.}~\bibnamefont {Le~Diz{\`e}s}}, \ and\
  \bibinfo {author} {\bibfnamefont {E.}~\bibnamefont {Villermaux}},\
  }\href@noop {} {\bibfield  {journal} {\bibinfo  {journal} {J. Fluid Mech.},\
  }\textbf {\bibinfo {volume} {577}},\ \bibinfo {pages} {241} (\bibinfo {year}
  {2007})}\BibitemShut {NoStop}%
\bibitem [{\citenamefont {Yang}\ \emph {et~al.}(2003)\citenamefont {Yang},
  \citenamefont {Prosperetti},\ and\ \citenamefont {Takagi}}]{Yang2003}%
  \BibitemOpen
  \bibfield  {author} {\bibinfo {author} {\bibfnamefont {B.}~\bibnamefont
  {Yang}}, \bibinfo {author} {\bibfnamefont {A.}~\bibnamefont {Prosperetti}}, \
  and\ \bibinfo {author} {\bibfnamefont {S.}~\bibnamefont {Takagi}},\
  }\href@noop {} {\bibfield  {journal} {\bibinfo  {journal} {Phys. Fluids},\
  }\textbf {\bibinfo {volume} {15}},\ \bibinfo {pages} {2640} (\bibinfo {year}
  {2003})}\BibitemShut {NoStop}%
\bibitem [{\citenamefont {Magnaudet}\ and\ \citenamefont
  {Eames}(2000)}]{Magnaudet2000}%
  \BibitemOpen
  \bibfield  {author} {\bibinfo {author} {\bibfnamefont {J.}~\bibnamefont
  {Magnaudet}}\ and\ \bibinfo {author} {\bibfnamefont {I.}~\bibnamefont
  {Eames}},\ }\href@noop {} {\bibfield  {journal} {\bibinfo  {journal} {Annu.
  Rev. Fluid Mech.},\ }\textbf {\bibinfo {volume} {32}},\ \bibinfo {pages}
  {659} (\bibinfo {year} {2000})}\BibitemShut {NoStop}%
\bibitem [{\citenamefont {Davies}\ and\ \citenamefont
  {Taylor}(1950)}]{Davies1950}%
  \BibitemOpen
  \bibfield  {author} {\bibinfo {author} {\bibfnamefont {R.~M.}\ \bibnamefont
  {Davies}}\ and\ \bibinfo {author} {\bibfnamefont {G.}~\bibnamefont
  {Taylor}},\ }\href@noop {} {\bibfield  {journal} {\bibinfo  {journal} {Proc.
  R. Soc. London A},\ }\textbf {\bibinfo {volume} {200}},\ \bibinfo {pages}
  {375} (\bibinfo {year} {1950})}\BibitemShut {NoStop}%
\bibitem [{\citenamefont {Walters}\ and\ \citenamefont
  {Davidson}(1963)}]{Walters1963}%
  \BibitemOpen
  \bibfield  {author} {\bibinfo {author} {\bibfnamefont {J.~K.}\ \bibnamefont
  {Walters}}\ and\ \bibinfo {author} {\bibfnamefont {J.~F.}\ \bibnamefont
  {Davidson}},\ }\href@noop {} {\bibfield  {journal} {\bibinfo  {journal} {J.
  Fluid Mech.},\ }\textbf {\bibinfo {volume} {17}},\ \bibinfo {pages} {321}
  (\bibinfo {year} {1963})}\BibitemShut {NoStop}%
\bibitem [{\citenamefont {Bonometti}\ and\ \citenamefont
  {Magnaudet}(2006)}]{Bonometti2006}%
  \BibitemOpen
  \bibfield  {author} {\bibinfo {author} {\bibfnamefont {T.}~\bibnamefont
  {Bonometti}}\ and\ \bibinfo {author} {\bibfnamefont {J.}~\bibnamefont
  {Magnaudet}},\ }\href@noop {} {\bibfield  {journal} {\bibinfo  {journal}
  {Phys. Fluids},\ }\textbf {\bibinfo {volume} {18}},\ \bibinfo {pages}
  {052102} (\bibinfo {year} {2006})}\BibitemShut {NoStop}%
\bibitem [{\citenamefont {Marten}\ \emph {et~al.}(1996)\citenamefont {Marten},
  \citenamefont {Shariff}, \citenamefont {Psarakos},\ and\ \citenamefont
  {White}}]{Marten1996}%
  \BibitemOpen
  \bibfield  {author} {\bibinfo {author} {\bibfnamefont {K.}~\bibnamefont
  {Marten}}, \bibinfo {author} {\bibfnamefont {K.}~\bibnamefont {Shariff}},
  \bibinfo {author} {\bibfnamefont {S.}~\bibnamefont {Psarakos}}, \ and\
  \bibinfo {author} {\bibfnamefont {D.~J.}\ \bibnamefont {White}},\ }\href@noop
  {} {\bibfield  {journal} {\bibinfo  {journal} {Sci. Am.},\ }\textbf {\bibinfo
  {volume} {275}},\ \bibinfo {pages} {82} (\bibinfo {year} {1996})}\BibitemShut
  {NoStop}%
\bibitem [{\citenamefont {Shelby}(2005)}]{Shelby2005}%
  \BibitemOpen
  \bibfield  {author} {\bibinfo {author} {\bibfnamefont {J.~E.}\ \bibnamefont
  {Shelby}},\ }\href@noop {} {\emph {\bibinfo {title} {Introduction to Glass
  Science and Technology}}},\ \bibinfo {edition} {2nd}\ ed.\ (\bibinfo
  {publisher} {The Royal Society of Chemistry},\ \bibinfo {year}
  {2005})\BibitemShut {NoStop}%
\bibitem [{\citenamefont {Gu{\'e}zennec}\ \emph {et~al.}(2005)\citenamefont
  {Gu{\'e}zennec}, \citenamefont {Huber}, \citenamefont {Patisson},
  \citenamefont {Sessiecq}, \citenamefont {Birat},\ and\ \citenamefont
  {Ablitzer}}]{Guezennec2005}%
  \BibitemOpen
  \bibfield  {author} {\bibinfo {author} {\bibfnamefont {A.-G.}\ \bibnamefont
  {Gu{\'e}zennec}}, \bibinfo {author} {\bibfnamefont {J.-C.}\ \bibnamefont
  {Huber}}, \bibinfo {author} {\bibfnamefont {F.}~\bibnamefont {Patisson}},
  \bibinfo {author} {\bibfnamefont {P.}~\bibnamefont {Sessiecq}}, \bibinfo
  {author} {\bibfnamefont {J.-P.}\ \bibnamefont {Birat}}, \ and\ \bibinfo
  {author} {\bibfnamefont {D.}~\bibnamefont {Ablitzer}},\ }\href@noop {}
  {\bibfield  {journal} {\bibinfo  {journal} {Powder Technol.},\ }\textbf
  {\bibinfo {volume} {157}},\ \bibinfo {pages} {2} (\bibinfo {year}
  {2005})}\BibitemShut {NoStop}%
\bibitem [{\citenamefont {Clift}\ \emph {et~al.}(1978)\citenamefont {Clift},
  \citenamefont {Grace},\ and\ \citenamefont {Weber}}]{Clift1978}%
  \BibitemOpen
  \bibfield  {author} {\bibinfo {author} {\bibfnamefont {R.}~\bibnamefont
  {Clift}}, \bibinfo {author} {\bibfnamefont {J.~R.}\ \bibnamefont {Grace}}, \
  and\ \bibinfo {author} {\bibfnamefont {M.~E.}\ \bibnamefont {Weber}},\
  }\href@noop {} {\emph {\bibinfo {title} {Bubbles, Drops, and Particles}}}\
  (\bibinfo  {publisher} {Academic Press},\ \bibinfo {year} {1978})\BibitemShut
  {NoStop}%
\bibitem [{\citenamefont {Lohse}\ \emph {et~al.}(2004)\citenamefont {Lohse},
  \citenamefont {Bergmann}, \citenamefont {Mikkelsen}, \citenamefont
  {Zeilstra}, \citenamefont {van~der Meer}, \citenamefont {Versluis},
  \citenamefont {van~der Weele}, \citenamefont {van~der Hoef},\ and\
  \citenamefont {Kuipers}}]{Lohse2004}%
  \BibitemOpen
  \bibfield  {author} {\bibinfo {author} {\bibfnamefont {D.}~\bibnamefont
  {Lohse}}, \bibinfo {author} {\bibfnamefont {R.}~\bibnamefont {Bergmann}},
  \bibinfo {author} {\bibfnamefont {R.}~\bibnamefont {Mikkelsen}}, \bibinfo
  {author} {\bibfnamefont {C.}~\bibnamefont {Zeilstra}}, \bibinfo {author}
  {\bibfnamefont {D.}~\bibnamefont {van~der Meer}}, \bibinfo {author}
  {\bibfnamefont {M.}~\bibnamefont {Versluis}}, \bibinfo {author}
  {\bibfnamefont {K.}~\bibnamefont {van~der Weele}}, \bibinfo {author}
  {\bibfnamefont {M.}~\bibnamefont {van~der Hoef}}, \ and\ \bibinfo {author}
  {\bibfnamefont {H.}~\bibnamefont {Kuipers}},\ }\href@noop {} {\bibfield
  {journal} {\bibinfo  {journal} {Phys. Rev. Lett.},\ }\textbf {\bibinfo
  {volume} {93}},\ \bibinfo {pages} {198003} (\bibinfo {year}
  {2004})}\BibitemShut {NoStop}%
\bibitem [{\citenamefont {Duclaux}\ \emph {et~al.}(2007)\citenamefont
  {Duclaux}, \citenamefont {Caill{\'e}}, \citenamefont {Duez}, \citenamefont
  {Ybert}, \citenamefont {Bocquet},\ and\ \citenamefont
  {Clanet}}]{Duclaux2007}%
  \BibitemOpen
  \bibfield  {author} {\bibinfo {author} {\bibfnamefont {V.}~\bibnamefont
  {Duclaux}}, \bibinfo {author} {\bibfnamefont {F.}~\bibnamefont {Caill{\'e}}},
  \bibinfo {author} {\bibfnamefont {C.}~\bibnamefont {Duez}}, \bibinfo {author}
  {\bibfnamefont {C.}~\bibnamefont {Ybert}}, \bibinfo {author} {\bibfnamefont
  {L.}~\bibnamefont {Bocquet}}, \ and\ \bibinfo {author} {\bibfnamefont
  {C.}~\bibnamefont {Clanet}},\ }\href@noop {} {\bibfield  {journal} {\bibinfo
  {journal} {J. Fluid Mech.},\ }\textbf {\bibinfo {volume} {591}},\ \bibinfo
  {pages} {1} (\bibinfo {year} {2007})}\BibitemShut {NoStop}%
\bibitem [{\citenamefont {Oguz}\ and\ \citenamefont
  {Prosperetti}(1993)}]{Oguz1993}%
  \BibitemOpen
  \bibfield  {author} {\bibinfo {author} {\bibfnamefont {H.~N.}\ \bibnamefont
  {Oguz}}\ and\ \bibinfo {author} {\bibfnamefont {A.}~\bibnamefont
  {Prosperetti}},\ }\href@noop {} {\bibfield  {journal} {\bibinfo  {journal}
  {J. Fluid Mech.},\ }\textbf {\bibinfo {volume} {257}},\ \bibinfo {pages}
  {111} (\bibinfo {year} {1993})}\BibitemShut {NoStop}%
\bibitem [{\citenamefont {Burton}\ \emph {et~al.}(2005)\citenamefont {Burton},
  \citenamefont {Waldrep},\ and\ \citenamefont {Taborek}}]{Burton2005}%
  \BibitemOpen
  \bibfield  {author} {\bibinfo {author} {\bibfnamefont {J.~C.}\ \bibnamefont
  {Burton}}, \bibinfo {author} {\bibfnamefont {R.}~\bibnamefont {Waldrep}}, \
  and\ \bibinfo {author} {\bibfnamefont {P.}~\bibnamefont {Taborek}},\
  }\href@noop {} {\bibfield  {journal} {\bibinfo  {journal} {Phys. Rev.
  Lett.},\ }\textbf {\bibinfo {volume} {94}},\ \bibinfo {pages} {184502}
  (\bibinfo {year} {2005})}\BibitemShut {NoStop}%
\bibitem [{Note1()}]{Note1}%
  \BibitemOpen
  \bibinfo {note} {In the context of the deep seal of a transient
  impact-generated cavity, the 2D Rayleigh-Plesset model is also in excellent
  agreement with the experiments, even if the vertical motions are neglected
  \protect \citep {Bergmann2009}. This suggests that the relevant mechanism for
  the pinch-off is radial in essence.}\BibitemShut {Stop}%
\bibitem [{\citenamefont {Bergmann}\ \emph {et~al.}(2008)\citenamefont
  {Bergmann}, \citenamefont {Dejong}, \citenamefont {Choimet}, \citenamefont
  {Van Der~Meer},\ and\ \citenamefont {Lohse}}]{Bergmann2008}%
  \BibitemOpen
  \bibfield  {author} {\bibinfo {author} {\bibfnamefont {R.}~\bibnamefont
  {Bergmann}}, \bibinfo {author} {\bibfnamefont {E.}~\bibnamefont {Dejong}},
  \bibinfo {author} {\bibfnamefont {J.-B.}\ \bibnamefont {Choimet}}, \bibinfo
  {author} {\bibfnamefont {D.}~\bibnamefont {Van Der~Meer}}, \ and\ \bibinfo
  {author} {\bibfnamefont {D.}~\bibnamefont {Lohse}},\ }\href@noop {}
  {\bibfield  {journal} {\bibinfo  {journal} {J. Fluid Mech.},\ }\textbf
  {\bibinfo {volume} {600}},\ \bibinfo {pages} {19} (\bibinfo {year}
  {2008})}\BibitemShut {NoStop}%
\bibitem [{\citenamefont {Whitham}(1974)}]{Whitham1974}%
  \BibitemOpen
  \bibfield  {author} {\bibinfo {author} {\bibfnamefont {G.~B.}\ \bibnamefont
  {Whitham}},\ }\href@noop {} {\emph {\bibinfo {title} {Linear and Nonlinear
  Waves}}}\ (\bibinfo  {publisher} {John Wiley \& Sons},\ \bibinfo {year}
  {1974})\BibitemShut {NoStop}%
\bibitem [{\citenamefont {Simpson}(1997)}]{Simpson1997}%
  \BibitemOpen
  \bibfield  {author} {\bibinfo {author} {\bibfnamefont {J.~E.}\ \bibnamefont
  {Simpson}},\ }\href@noop {} {\emph {\bibinfo {title} {Gravity Currents in the
  Environment and the Laboratory}}},\ \bibinfo {edition} {2nd}\ ed.\ (\bibinfo
  {publisher} {Cambridge University Press},\ \bibinfo {year}
  {1997})\BibitemShut {NoStop}%
\bibitem [{\citenamefont {Bolanos-Jimenez}\ \emph {et~al.}(2008)\citenamefont
  {Bolanos-Jimenez}, \citenamefont {Sevilla}, \citenamefont {Martinez-Bazan},\
  and\ \citenamefont {Gordillo}}]{Bolanos-Jimenez2008}%
  \BibitemOpen
  \bibfield  {author} {\bibinfo {author} {\bibfnamefont {R.}~\bibnamefont
  {Bolanos-Jimenez}}, \bibinfo {author} {\bibfnamefont {A.}~\bibnamefont
  {Sevilla}}, \bibinfo {author} {\bibfnamefont {C.}~\bibnamefont
  {Martinez-Bazan}}, \ and\ \bibinfo {author} {\bibfnamefont {J.~M.}\
  \bibnamefont {Gordillo}},\ }\href@noop {} {\bibfield  {journal} {\bibinfo
  {journal} {Phys. Fluids},\ }\textbf {\bibinfo {volume} {20}},\ \bibinfo
  {pages} {112104} (\bibinfo {year} {2008})}\BibitemShut {NoStop}%
\bibitem [{\citenamefont {Bartolo}\ \emph {et~al.}(2006)\citenamefont
  {Bartolo}, \citenamefont {Josserand},\ and\ \citenamefont
  {Bonn}}]{Bartolo2006}%
  \BibitemOpen
  \bibfield  {author} {\bibinfo {author} {\bibfnamefont {D.}~\bibnamefont
  {Bartolo}}, \bibinfo {author} {\bibfnamefont {C.}~\bibnamefont {Josserand}},
  \ and\ \bibinfo {author} {\bibfnamefont {D.}~\bibnamefont {Bonn}},\
  }\href@noop {} {\bibfield  {journal} {\bibinfo  {journal} {Phys. Rev.
  Lett.},\ }\textbf {\bibinfo {volume} {96}},\ \bibinfo {pages} {124501}
  (\bibinfo {year} {2006})}\BibitemShut {NoStop}%
\bibitem [{\citenamefont {Gekle}\ and\ \citenamefont
  {Gordillo}(2010)}]{Gekle2010}%
  \BibitemOpen
  \bibfield  {author} {\bibinfo {author} {\bibfnamefont {S.}~\bibnamefont
  {Gekle}}\ and\ \bibinfo {author} {\bibfnamefont {J.~M.}\ \bibnamefont
  {Gordillo}},\ }\href@noop {} {\bibfield  {journal} {\bibinfo  {journal} {J.
  Fluid Mech.},\ }\textbf {\bibinfo {volume} {663}},\ \bibinfo {pages} {293}
  (\bibinfo {year} {2010})}\BibitemShut {NoStop}%
\bibitem [{\citenamefont {Bergmann}\ \emph {et~al.}(2009)\citenamefont
  {Bergmann}, \citenamefont {Van Der~Meer}, \citenamefont {Gekle},
  \citenamefont {Van Der~Bos},\ and\ \citenamefont {Lohse}}]{Bergmann2009}%
  \BibitemOpen
  \bibfield  {author} {\bibinfo {author} {\bibfnamefont {R.}~\bibnamefont
  {Bergmann}}, \bibinfo {author} {\bibfnamefont {D.}~\bibnamefont {Van
  Der~Meer}}, \bibinfo {author} {\bibfnamefont {S.}~\bibnamefont {Gekle}},
  \bibinfo {author} {\bibfnamefont {A.}~\bibnamefont {Van Der~Bos}}, \ and\
  \bibinfo {author} {\bibfnamefont {D.}~\bibnamefont {Lohse}},\ }\href@noop {}
  {\bibfield  {journal} {\bibinfo  {journal} {J. Fluid Mech.},\ }\textbf
  {\bibinfo {volume} {633}},\ \bibinfo {pages} {381} (\bibinfo {year}
  {2009})}\BibitemShut {NoStop}%
\end{thebibliography}%

\end{document}